\documentclass[pre,twocolumn,unsortedaddress,floatfix,amssymb,showpacs]{revtex4-1}
\usepackage{times}
\usepackage{graphicx}
\usepackage{amsmath}
\usepackage{color}
\usepackage{natbib}
\bibpunct{[}{]}{,}{n}{}{;}

\graphicspath{{gfx_article/}} 
\DeclareGraphicsExtensions{.eps, .eps.bz2}

\newcommand{\br}{\textbf{r}}

\begin{document}

\title{Event distributions of polymer translocation}

\author{R. P. Linna}
\author{K. Kaski}
\affiliation{
  Department of Biomedical Engineering and Computational Science,
  Aalto University, P.O. Box 12200, FI-00076 Aalto, Finland
  }


\begin{abstract}
We present event distributions for the polymer translocation obtained by extensive Langevin dynamics simulations. 
Such distributions have not been reported previously and they provide new understanding of the stochastic 
characteristics of the process. We extract at a high length 
scale resolution distributions of polymer segments that continuously traverse through a nanoscale pore. 
The obtained log-normal distributions together with the characteristics of polymer translocation suggest that it is describable
as a multiplicative stochastic process. In spite of its clear out-of-equilibrium nature the forced 
translocation is surprisingly similar to the unforced case. We find forms for the distributions almost unaltered 
with a common cut-off length. We show that the individual short-segment and short-time movements inside the pore give 
the scaling relations $\tau \sim N^\alpha$ and $\tau \sim f^{-\beta}$ for the polymer translocation. 
\end{abstract}

\maketitle

\section{Introduction}

Polymer translocation through a nano-scale pore has been under intensive research 
since the experimental study by Kasianowicz {\it et al.}~\cite{kasianowicz} and 
the first theoretical treatment by Sung and Park~\cite{sung} and the related study by Muthukumar~\cite{muthukumar}. The 
important treatment in~\cite{chuang,kantor} set the valid limits for the scaling exponents obtained for the unforced and 
forced translocation. The highly non-equilibrium nature of the process was captured in the first 
analytical treatment based on the tension propagation present in the forced translocation by 
Sakaue~\cite{sakaue}. Broadly speaking, the present theoretical translocation research is spanned 
between these few cornerstones.

We have previously shown that the forced polymer translocation 
takes place out of equilibrium and is mainly determined by the balance between the 
driving pore force and the drag force exerted on the polymer on the {\it cis} side~\cite{ourepl,ourpre}. 
The drag force magnitude was shown to change with the tension propagation, which shows as an increase in the number 
of moving polymer beads $N_m$ on the {\it cis} side. The drag force was shown to consequently decrease with $N_m$ as 
the polymer translocates to the {\it trans} side. The polymer crowding on the {\it trans} side was shown to modify 
this dynamics and be enhanced by increasing the pore force. The crowding  in turn increases  the scaling 
exponent $\alpha$ in $\tau \sim N^\alpha$ where $\tau$ is the translocation time and $N$ the number of polymer beads, 
{\it i.e.} the polymer length~\cite{ourepl,ourpre}. We also showed that $\alpha$ decreases with increasing the pore 
friction~\cite{ourpre2}. Our description of the process  is very closely aligned with Sakaue's 
analytical treatment that has been adopted and expanded {\it e.g.} in~\cite{dubbeldam2011}. Sakaue's treatment and 
the characteristics described above have been incorporated in a coarse-grained model in~\cite{ikonen}.

The coarse-grained description of the forced polymer translocation outlined above says very little about the stochastic 
nature of the process. The force balance and tension propagation fall within the realm of classical mechanics. Although 
they give the correct average characteristics of the forced polymer translocation, for example the question of how 
frequently the translocating polymer stops and reverses its direction and how long segments typically pass through the 
pore ``in one go'' and how long such transitions take remain unanswered. Questions such as these have relevance for implementing DNA sequencing. The force balance and tension propagation framework can outline the force 
dependence of the observed scaling of the translocation time $\tau$ with the polymer length $N$ but cannot describe 
how this scaling comes about in the first place. In this paper our motivation is to characterize the stochastic nature of the 
polymer translocation and the origin of the scaling by collecting data on the individual transitions of polymer segments.

A fundamental question is
how the forced translocation differs stochastically from the unforced process. The forced polymer translocation is a highly 
non-equilibrium process even for a modest driving pore force. From the pore outward on both the {\it cis} and the {\it trans} 
side an increasing portion of the polymer is lifted out of thermal equilibrium in the course of 
translocation~\cite{ourepl,ourpre,ourunf}. The correlations along the polymer contour increase due to the polymer being driven 
out of equilibrium. Hence, it is natural to expect that applying the pore force may alter the 
obtained distributions describing the transfer of polymer segments through the pore. As we will show, these distributions also 
shed light on the way the observed scaling relations emerge. We have previously found that the waiting-time distribution, 
{\it i.e.} the numbers of transitions $n$ that take a time $\Delta t$ plotted as a function of $\Delta t$, $n(\Delta t)$, obey closely the 
Poisson distribution~\cite{ourunf}. The events, {\it i.e.} the transitions of 
individual segments were registered with the resolution of a polymer bond length $b$. It is clear that the polymer translocation 
for which the translocation time $\tau$ scales with the polymer length $N$ cannot be a Poissonian process. One is then naturally 
led to presume that the polymer translocation shows additional correlations if observed at smaller length scales than $b$.

The outline of the paper is as follows: In Section~\ref{cm} the computational model and its relation to experiments is 
explained. In Subsection~\ref{ne} we define the events that we register from the simulated polymer translocations. The 
multiplicative stochastic process is defined and the polymer translocation as such a process described in Subsection~\ref{ln}. 
The results are reported and discussed in the remaining parts of Section~\ref{rs}. Finally, in Section~\ref{cn} we summarize the 
main conclusions based on our findings.

\section{ The Computational Model}
\label{cm}

\subsection{The polymer model}

We use the standard bead-spring chain as a coarse-grained model for the polymer. Here, 
adjacent monomers are connected with anharmonic springs, described by the  finitely
extensible nonlinear elastic (FENE) potential 
\begin{equation}
\label{fene}
U_F = - \frac{K}{2} R^2 \ln \big ( 1- \frac{r^2}{R^2} \big ),
\end{equation}
where $r$ is the length of an effective bond and $R = 1.5 \sigma$ is the maximum bond
length. The shifted Lennard-Jones (LJ) potential
\begin{equation}
\label{lj}
U_{LJ} = 4 \epsilon \left[ \left(\frac{\sigma}{r}\right)^{12} -
\left(\frac{\sigma}{r}\right)^{6} + \frac{1}{4} \right], \: r \leq 2^{1/6},
\end{equation}
is applied between all beads of distance $r$ apart. The parameter values
were chosen as $\epsilon = 1.0$, $\sigma = 1.0$ and $K = 30 / \sigma^2$. As there is no 
attractive part in the used LJ potential, the model is for a polymer in a good solvent.
We apply no bending potential, rendering the polymer as a freely-jointed chain (FJC), 
instead of a worm-like chain (WLC).  We have previously shown that the 
difference of the two models when simulating 
polymer elasticity in mechanical stretching or in flow is insignificant~\cite{linna}.

\subsection{The dynamics}
\label{dn}

The dynamics of the polymer translocation was simulated using Ermak's implementation of 
Brownian dynamics~\cite{ermak}. Accordingly, the time derivative of the momentum of the
polymer bead $i$ is given by
\begin{equation}
\label{brownian}
\dot{\mathbf{p}}_i(t) = - \xi \mathbf{p}_i(t) + \eta_i(t) + f(\br_i),
\end{equation}
where $\xi$, $\mathbf{p}_i(t)$, $\eta_i(t)$, and $f(\br_i)$ are the 
friction constant, momentum, random force of the bead $i$, and
external driving force, respectively. When applied, $f(\br_i)$ is constant
and exerted only inside the pore. Velocity Verlet was applied in the time 
integration~\cite{vangunsteren}. In the present study the parameter values in reduced 
units~\cite{allen} were as follows: $\xi = 0.5$ and $\eta_i(t)$ is related to $\xi$ by 
the fluctuation-dissipation theorem. The mass of a polymer bead is $m = 16$. The time 
step used in the numerical integration of equations of motion was $\delta t = 0.001$. 
Partly due to the out-of- equilibrium nature of driven 
translocation, changing either $\xi$ or $m$ changes the obtained scaling of the 
translocation time with the number of beads, $\tau \sim N^\alpha$~\cite{ourpre2}.

\subsection{The wall and the pore}
\label{wp}

The wall comprises two aligned surfaces the distance $l = 3b$ apart, where $b = 1$ is the Kuhn length for the model polymer. 
$l$ is thus the length of the pore.  No-slip boundary conditions applied on the polymer beads hitting the surfaces prevent 
them from entering the wall. The pore is modelled as a cylindrical potential whose center axis is perpendicular to 
the wall and extends through it. The pore diameter is $1.2 \sigma$. Inside the pore, the  
cylindrically symmetric damped harmonic potential $U_h$ pulls the beads toward the pore axis with force 
\begin{equation}
f_h = -\nabla U_h = -k r_p - c v_p, 
\end{equation}
where $k = 100$, $c = 1$, $r_p$ is the distance of a polymer bead from the pore axis, and $v_p$ is its velocity component 
perpendicular to the axis. As mentioned in Subsection~\ref{dn}, the pore force $f(\br_i)$ in 
Eq.~(\ref{brownian}) is exerted 
only inside the pore in the case of forced translocation. To prevent the pore force from fluctuating with 
the changing number ($2$ or $3$) of polymer beads inside the pore we calculate the exact fraction of the 
polymer inside the pore and adjust the force  magnitude applied on each bead inside the pore so that the pore 
exerts a constant force per polymer length.

There is no explicit adhesion between the polymer and the pore in 
our model. While adhesion is of importance in the DNA translocation experiments, see {\it e.g.}~\cite{wanunu}, here we want to 
characterize the basic translocation process prior to embarking on the effects of relevant additional interactions. The 
transitions and pauses of the polymer segments inside the pore hence result from the changes in the force exerted on these 
segments inside the pore. These changes in the force, in turn, reflect the conformational changes of the polymer segments 
outside the pore and their changing interactions with the Brownian heat baths on the {\it cis} and {\it trans} sides of the wall.

\subsection{Relation to experiments}

In order to relate the computational force to a physical force inside a pore in experiments we 
need to relate the energy and length scale in our model to the physical energy and 
length scale. In our reduced units $kT = 1$ corresponds to $k_B \tilde{T}$, where $k_B$ is 
the Boltzmann constant, and the physical temperature $\tilde{T}$ is taken to be 
$300\ \text{K}$. The correspondence between computational and physical length scales 
can be established by taking the polymer bond length $b$ as the Kuhn length for the 
physical polymer. In SI units the bond length for our FJC model polymer can be 
obtained as $\tilde{b}=2\lambda_p$, where $\lambda_p$ is the persistence 
length, $40\ \text \AA$ for a ssDNA~\cite{tinland}.
The pore force per bond length in SI units, $\tilde{f}_{tot}$, is then obtained from the 
dimensionless pore force per bond segment, $f$, by relating 
$\tilde{f} \tilde{b}/k_B \tilde{T} = fb/kT$.  The effective pore force per bond of 
$f = 1$ thus corresponds to $\tilde{{f}} = 0.52\ \text{pN}$ per Kuhn length for a ssDNA. 
Since $l = 3b$, $f = 1$ corresponds to the total pore force 
$\tilde{{f}}_{tot} = 1.56\ \text{pN}$. If one uses Manning condensation leading to drastic charge 
reduction~\cite{meller,sauerbudge} to relate this to experiments, in the $\alpha$-HL pore a 
typical pore potential of $\approx 120\ \text{mV}$ would correspond to $\tilde{f}_{tot} \approx 5\ \text{pN}$. 
In the light of recent computer simulations the reduction of the effective force inside the $\alpha$-HL pore is not 
predominantly caused by the Manning condensation but by electro-osmotic flow that is driven by the motion of 
counterions along the surface of DNA~\cite{luan,vandorp}. However, experiments indicate that the estimate based 
on Manning condensation is in the right order of magnitude~\cite{keyser}.

\section{Results: Events Observed with High Resolution}
\label{rs}

Traditionally, an event in a translocation simulation is defined as a change of the polymer 
bead at a reference position, here the pore opening on the {\it trans} side. In order to discern 
correlations at a finer scale than $b$ we have implemented in our model 
the possibility to register polymer segment motion inside the pore with the resolution $b/10$. In 
addition, we redefine the event as a polymer segment traversed in one direction inside the pore.  
The event-related distributions were extracted from at least $1000$ individual 
translocation runs for each pore force. The runs ending in the polymers translocating 
to the {\it trans} side and the ones where they slided back to the {\it cis} side were 
identified. Distributions were determined for all runs and the {\it trans} and 
{\it cis} cases, separately.

\subsection{Numbers of events vs traversed segment lengths}
\label{ne}

First, we define the event as a polymer segment $\Delta s$ that traverses in either direction inside 
the pore without pausing. We sample these events at constant time intervals $t_{int}$. Already using 
this somewhat limited definition for an event reveals a log-normal distribution for the number of 
events $n_E$ {\it vs} $\Delta s$. In Fig.~\ref{segment_distr}(a) we show distributions $n_E(\Delta s)$ 
for different pore force magnitudes. The distributions for motion toward {\it cis} and {\it trans} are 
identical for unforced translocation. Increasing the pore force $f$ these distributions deviate due to the 
increasing proportion of events toward {\it trans}.

The total time $\tau$ it takes a polymer to translocate consists of forward and backward motion 
and pauses. In order to find how the scaling law $\tau \sim N^\alpha$ emerges we have to 
include the pauses in the definition of the events and register events without the time limit 
introduced by the time interval $t_{int}$. Accordingly, we define an event to be 
terminated only when the direction of the motion is reversed. The characteristics are the same as with 
the first event definition, only the distributions become wider. To compare the forms of the distributions 
we normalize them with the maximum number of events, $\hat{n}_e = n_e/n_{max}$. These normalized 
log-binned distributions are shown in Fig.~\ref{segment_distr}(b). For clarity, the separated 
distributions toward {\it cis} are shown only for the events defined in the first way.

\subsection{Log-normal distribution and multiplicative stochastic processes}
\label{ln}

The obtained log-normal distributions suggest that the polymer translocation is largely dominated by the 
underlying {\it multiplicative} stochastic process. For such a process the probability of an event 
$P^{(m)}_r$ composed of the succession of $m$ independent events with the probabilities 
$p_i$ ($1 \le i \le m$) is given by the 
product of these independent random variables $P^{(m)}_r = \prod_{i=1}^m p_i$, so that 
$\log P^{(m)}_r = \sum_{i=1}^m \log p_i$. With $m$ large, $\log P^{(m)}_r$ becomes a normal and, 
hence, $P^{(m)}_r$ a log-normal distribution due to the central limit theorem ~\cite{matsushita}.

In the case of polymer translocation this multiplicativity arises naturally, since we observe traversed 
polymer segments $\Delta s_n$. Here, $n$ indexes the registered segments. The traversed segments $\Delta s_n$ 
toward {\it trans} result from $m$ consecutive transitions of distance $s_i$ toward {\it trans}. An 
analogous statement applies, of course, for transitions toward {\it cis}.
Denoting the probabilities of these individual transitions as $p_i$ ($1 \le i \le m$), we see 
that the probability of $m$ such consecutive transitions is $P^{(m)}_r = \prod_{i=1}^m p_i$. Hence, the polymer 
translocation is seen to be of the simplest type of a multiplicative process. Since the 
traversed segments $\Delta s_n = m s_i$, the probability distribution $P(\Delta s_n)$ is of the same form as 
$P^{(m)}_r$. Taking the continuum limit, the enumerable set $\{\Delta s_n\}$ is replaced by the continuous 
variable $\Delta s$. In this limit $P(\Delta s)$ is log-normal due to the central limit theorem. In what follows 
we drop the subscript $n$ and use the symbol $\Delta s$ to denote the lengths of the traversed segments 
registered with resolution $b/10$.

There are subtleties in applying the central limit theorem to multiplicative processes, see~\cite{redner}. These 
are related to the effect of rare events. In our 
case these correspond to large $m$, or $\Delta s$. As we will see, the translocation is dominated by short 
$\Delta s$ events, so applying the central limit theorem is valid and the obtained log-normal distributions have 
the natural explanation described above.

The probabilities $p_i$ appear independently but are dependent on the stage of the translocation 
process. Transition toward {\it trans} gets more probable toward the end of the translocation process 
when the majority of the polymer is already translocated. Also the crowding of the polymer segments at 
large pore force changes this probability by different amounts at different stages of 
translocation~\cite{ourepl}. This differing of $p_i$'s at different stages of the process is 
characteristic for multiplicative stochastic processes, see {\it e.g.}~\cite{schenzle}. Due to the 
additional correlations in fluctuations multiplicative stochastic processes are typically highly 
non-equilibrium processes~\cite{schenzle,havlin}, which we have shown to be the case for the polymer 
translocation~\cite{ourepl,ourpre}.

The observed 
traversed polymer segments $\Delta s$ consist of successive individual transitions $s_i$ measured with 
resolution $b/10$. The transitions were seen to be slow enough that the registered $s_i = \pm b/10$. We obtain 
distributions of the log-normal form
\begin{equation}
P(\Delta s) \sim \frac{1}{\Delta s} \exp [- \ln(\Delta s/\Delta s_0)^2/2 \sigma^2],
\end{equation}
 where 
$\Delta s_0$ is the characteristic scale and $\sigma$ the standard deviation of the variable 
$\ln \Delta s$, see {\it e.g.}~\cite{sornette}. To further confirm that we are indeed sampling events from 
a distribution resulting from a multiplicative stochastic process we also registered each successive 
transition at times $i$ and collected distributions such that even successive transitions of different 
lengths in the same direction were identified as separate events. Again, clean log-normal distributions 
resulted.

The forms of the {\it trans} distributions are seen to become wider with increasing pore force $f$. 
This is due to the increased proportion of long-segment motion at larger $f$. 
In accordance, the {\it cis} distributions become narrower with increasing $f$. However, the 
distributions vary fairly little with $f$. Especially surprising is the small difference between the 
distributions for unforced and forced translocation in spite of the fact that forced translocation 
already at weak pore force was seen to be governed by a non-equilibrium force balance 
condition~\cite{ourepl,ourpre}. In what follows, we show results for distributions obtained using the 
latter event definition.

\subsection{Numbers of events vs event times}

\begin{figure}
\centerline{
\includegraphics[angle=0, width=0.23\textwidth]{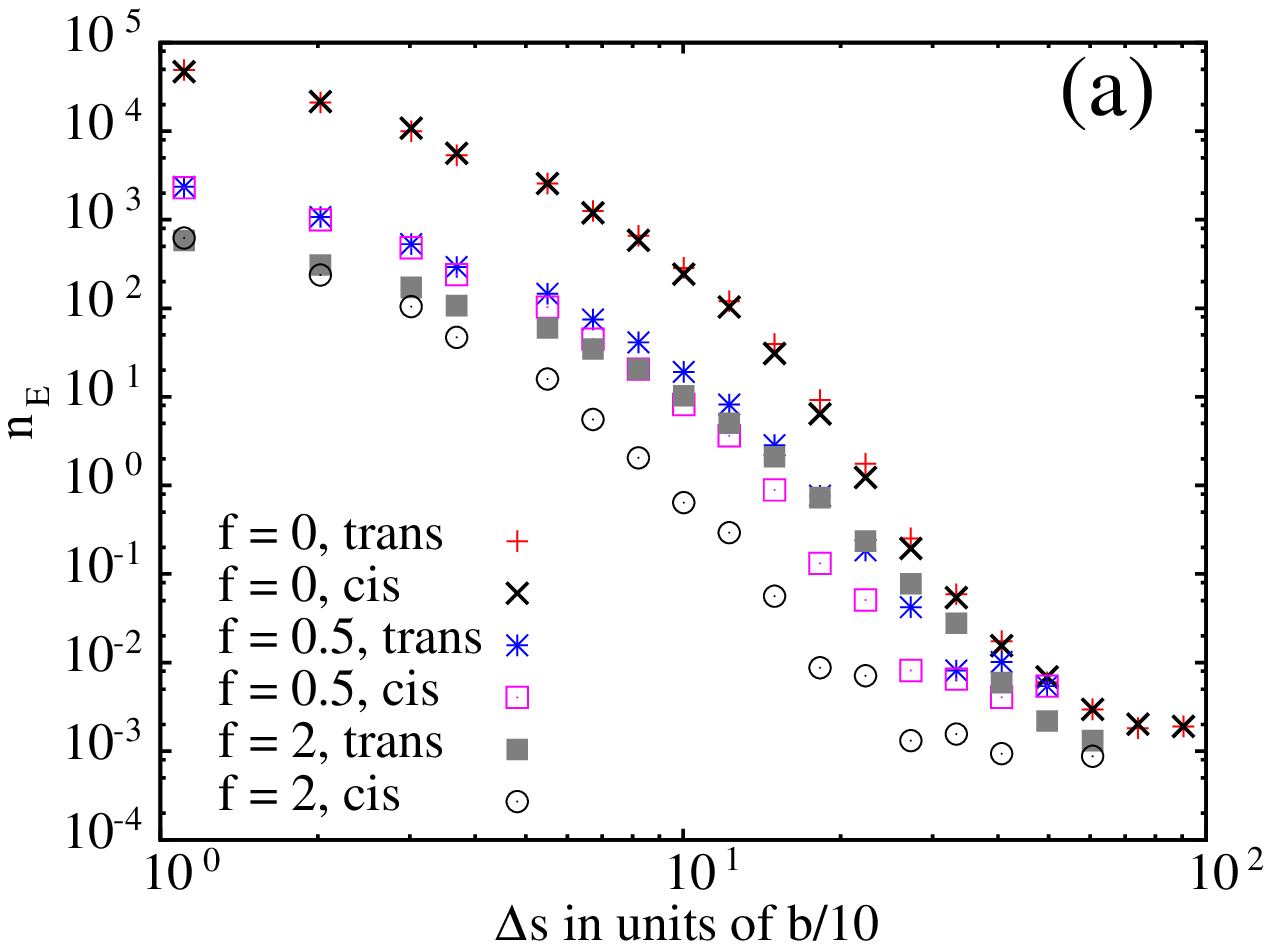}
\includegraphics[angle=0, width=0.23\textwidth]{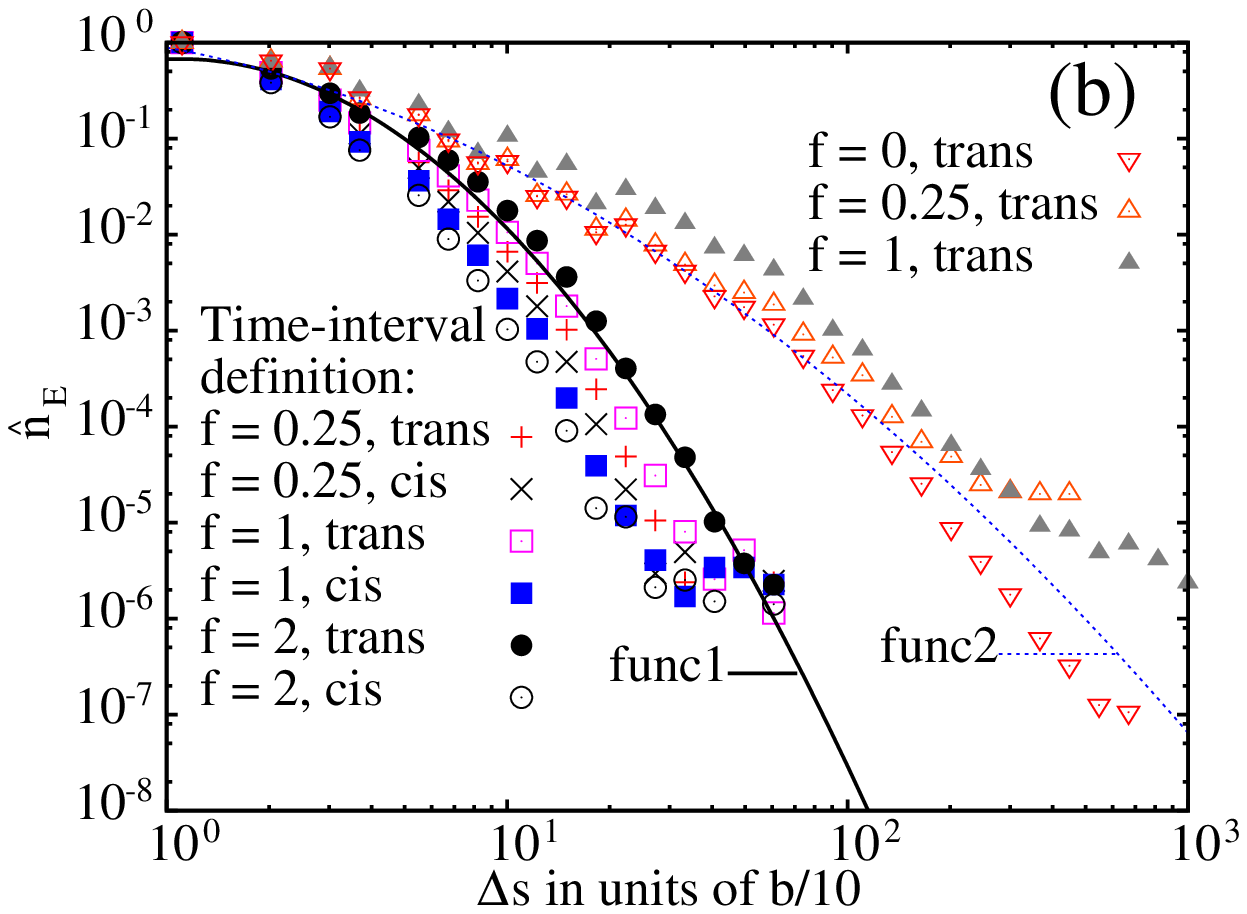}
}
\caption{
(Color online) Log-binned distributions of events assorted by the lengths of polymer 
segments that 
traverse in one direction, {\it trans} or {\it cis}, without pausing and sampled at time intervals, or 
without reversing for different pore force magnitudes $f$. The length of the polymer for all curves is 
$N = 200$. (a) The number of 
events $n_E$ vs. segment length $\Delta s$. $n_E$ is an average of numbers of events over several 
polymer translocations. (b) The numbers of events normalized with the maximum, $\hat{n}_E = n_E/n_{max}$
 to compare the distributions. All distributions follow closely log-normal forms. The log-normal 
functions plotted to guide the eye are 
$\textrm{func1} = 1/\Delta s *\exp(-\log(\Delta s/2)^2/1.2)$ and $\textrm{func2} = 1/\Delta s *\exp(-\log(\Delta s/2)^2/4$. The 
apparent power-law decay at $\Delta s > 300$ is an artefact due to the log-binning done on very 
few points there.
%
}
\label{segment_distr}
\end{figure}

\begin{figure}
\centerline{
\includegraphics[angle=0, width=0.4\textwidth]{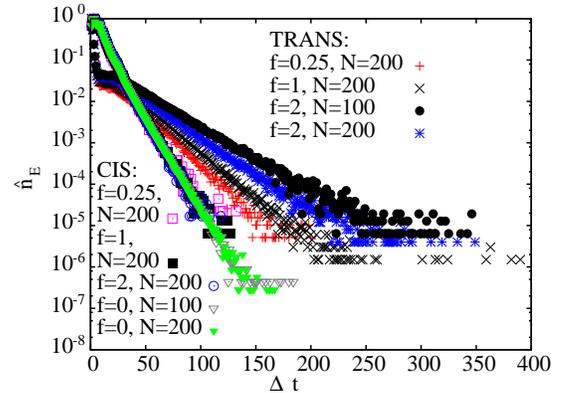}
}
\caption{
(Color online) The event-time distributions $n_E$ normalized with the maximum number of events. Semi-log scale shows the 
exponential decays. See the text for details.
}
\label{event_times}
\end{figure}

Next, we look at the distribution of events in terms of event times $\Delta t$. 
Fig.~\ref{event_times} 
shows the event-time distributions of the segments traversed toward {\it trans} and {\it cis} for 
$f = 1$ and $N \in \{50,100,200\}$. The number of events toward {\it trans} decays as 
$n_E \sim e^{-A \Delta t}$. The transfer times of the events toward {\it cis} obey 
$n_E \sim e^{-B \Delta t}$ for all $\Delta t$. $B \approx 0.11$ for  all $f$, whereas $A = 0.055$, 
$0.05$, and $0.043$ for $f = 0.25$, $1$, and $2$, respectively, when $N = 200$. So, the proportion of 
events taking a longer time increases with $f$, which is in keeping with Fig.~\ref{segment_distr}(b) 
where the proportion of long-segment events are seen to increase with $f$. This explicitly shows 
the weak increase of the characteristic correlation length with increasing $f$. The exponential decay 
$n_E \sim e^{-A \Delta t}$ is slightly slower for polymers of length $N = 100$ than of $N = 200$, see 
the plots for $f = 2$ in Fig.~\ref{event_times}. The dependence of $A$ on $N$ even for the maximum 
pore force $f =2$ is weak compared with its dependence on $f$ and hence is not expected to have any 
effect on the scaling $\tau \sim N^\alpha$. The maximum event times and 
segment lengths are seen to increase only weakly with $f$. In summary, the form of the distribution of 
events with event time varies slightly with $f$ and $N$ for events toward {\it trans} but stays unaltered 
for events toward {\it cis}.

\subsection{Distributions for constant segment lengths}

\begin{figure}
\centerline{
\includegraphics[angle=0, width=0.23\textwidth]{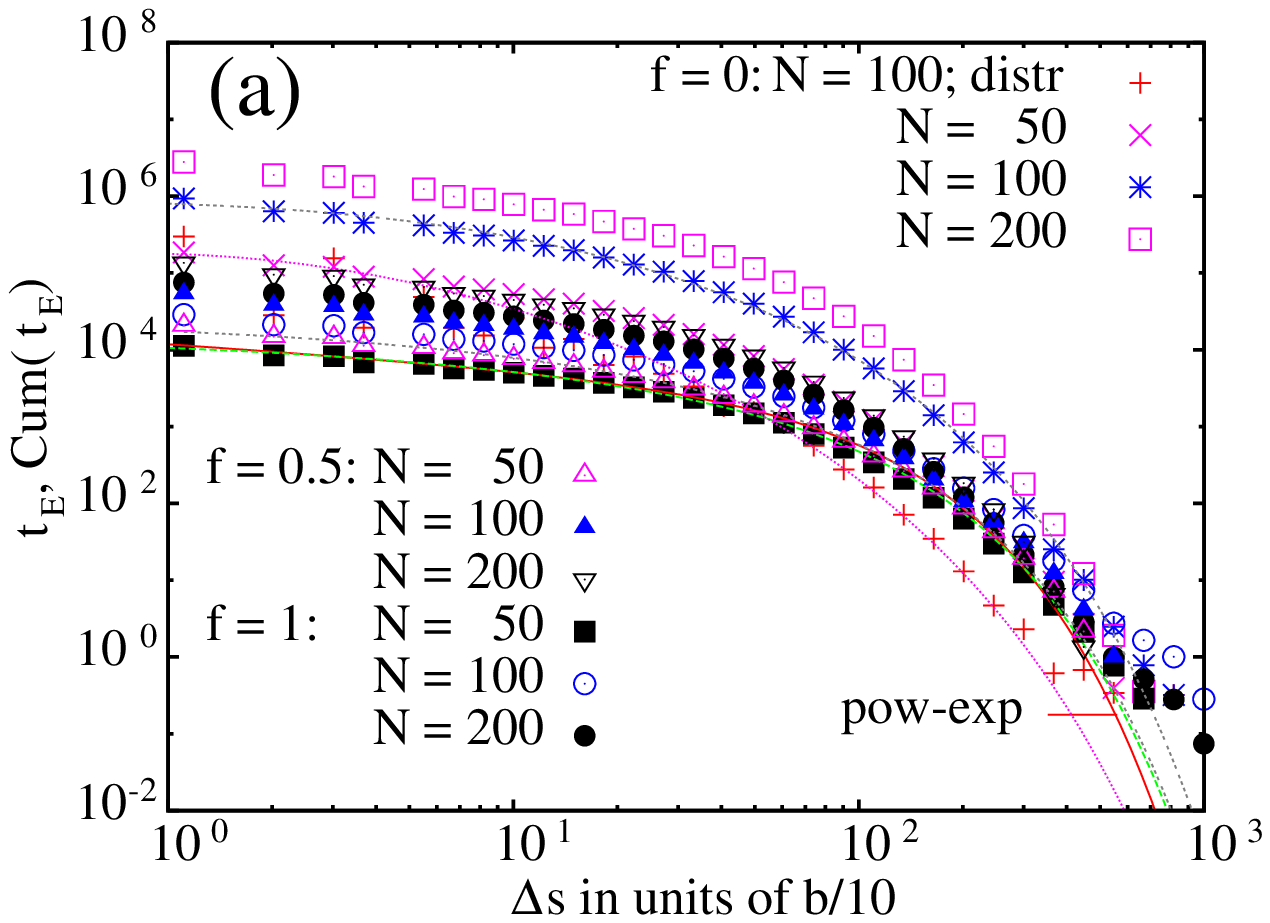}
\includegraphics[angle=0, width=0.23\textwidth]{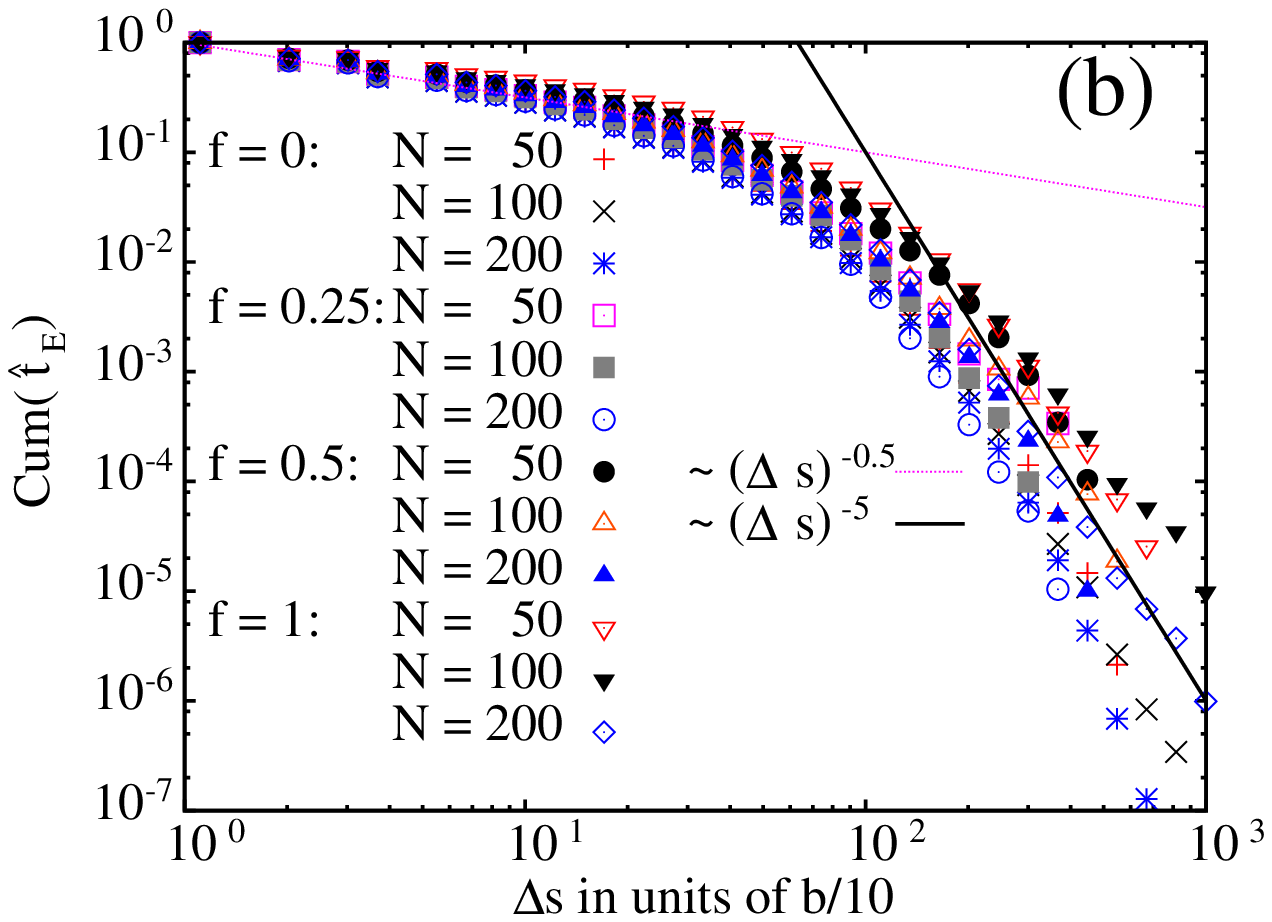}
}
\caption{
(Color online) (a) The distribution (distr) of event times $\Delta t$ with segment lengths $\Delta s$ 
for $f = 0$ and $N = 100$. The rest are cumulative distributions for $f = 0, 0.5$ and $1$, 
$N = 50$, $100$, and $200$. The cut-off segment length $\Delta s_c = 95$ and the characteristic 
scale $\Delta s_0 = 6$ for all fit functions. The dispersion $\sigma = 2$, $7$,and $9.5$ for $f = 0$,
 $0.5$, and $1$, respectively. Also shown is a fit with the function 
$\textrm{pow-exp} \sim \Delta s^{-0.3}\exp(-\Delta s/60)$ for the cumulative distribution for $f = 1$ 
and $N = 50$. (b) The cumulative distributions of (a) normalized by the maximum value. Lines 
$\sim (\Delta s)^{-0.5}$ and $\sim (\Delta s)^{-5}$ are drawn to guide the eye.
}
\label{cumsl}
\end{figure}

For DNA sequencing the important information is the time $\Delta t$ it takes a polymer to traverse a 
distance $\Delta s$ through the pore. Accordingly, we define $t_E(\Delta s)$ as the total elapsed time for transitions 
of constant distance $\Delta s$. $t_E$ consists of multiple separate events that take slightly different times $\Delta t$. 
We denote the number of such events $n_E(\Delta s_{\textrm{const}}, \Delta t)$. The total time spent on transferring segments of 
length $\Delta s_{\textrm{const}}$ is thus  
$t_E = \sum n_E(\Delta s_{\textrm{const}}, \Delta t) \cdot \Delta t(\Delta s_{\textrm{const}})$, where the sum runs over all 
registered $\Delta t$ for $\Delta s_{\textrm{const}}$. In 
Fig.~\ref{cumsl} we show $t_E$ for all registered discrete constant values of $\Delta s$. In what follows we omit the sub text 
'const' from $\Delta s$. The obtained distributions are of the form
\begin{equation}
P(t_E(\Delta s)) \sim \frac{1}{\Delta s} \exp [- \ln(\Delta s/\Delta s_0)^2/2 \sigma^2] 
\exp(- \Delta s/ \Delta s_c),
\end{equation}
where $\Delta s_c$ is a finite cut-off. All these distributions can be fit by keeping 
$\Delta s_c$ and $\Delta s_0$ constant and varying $\sigma$ in the log-normal part. The distribution data show 
fluctuations at small $\Delta s$, so we show cumulative distributions for improved quality. For a continuous variable $x$ the 
cumulative distribution $CP$ of the original distribution $P$ is defined as $CP(x) = \int_{+\infty}^0 P(x) dx$. So, for the 
discrete variable $\Delta s$ we have $CP(t_E(\Delta s)) = \sum_{\Delta_s = {+\infty}}^0 P(t_E(\Delta s))$. Naturally, all these 
distributions can be fit with the cumulative form 
\begin{equation}
CP(t_E(\Delta s)) \sim 1 - \textrm{erf}\lbrack \sqrt{\ln(\Delta s/\Delta s_0)^2/2 \sigma^2 + \Delta s/ 
\Delta s_c} \rbrack. 
\end{equation}
Again, the fits can be made by keeping $\Delta s_c$ and $\Delta s_0$ constant and varying $\sigma$, 
which increases with the pore force $f$. Unlike for the unforced translocation, for the forced 
translocation the broader cumulative log-normal distributions can also be fit with a 
decreasing power law and an exponential cutoff, $\Delta s^{-\beta}\exp(-\Delta s/\Delta s_{co})$.

The translocation times can be read off from the maximum values of the cumulative distributions. 
Statistics is not as good as when $\tau$ is directly measured from and averaged over individual 
simulations. The exponent values $\alpha$ obtained this way are $\alpha \gtrapprox 2.2$ for $f = 0$, and 
$\alpha \approx 1.4, 1.4, 1.45$ for $f = 0.25, 0.5, 1$, respectively. These approximative values 
are in keeping with our previous results and reported values for $\alpha$ in general. The same values for 
$\alpha$ are obtained by direct measurements of $\tau$, so this also verifies that the distributions were 
obtained correctly. The dependence of $\alpha$ on $f$ is mild compared to some of our earlier simulations. 
This may be caused by our more refined way of taking care that the pore force stays strictly constant, see 
Subsection~\ref{wp}. The effect of fluctuating pore force will be investigated in more detail in a forthcoming 
publication.

It is noteworthy that for the unforced and forced translocation there is a common constant cutoff 
length $\Delta s_{co} \approx 95$ that does not increase with the pore force or change with the polymer 
length $N$. The proportion of traversed longer segments, however, 
increases with $f$, which is seen as a broadening of the log-normal distribution. Again, the form of 
the distribution changes very little with $f$ or $N$, which can be more clearly seen from the 
normalized distributions in Fig~\ref{cumsl}(b). It would be tempting to think that the scaling 
$\tau \sim N^\alpha$ resulted from the broad log-normal distribution of event times 
$t_E(\Delta s)$. However, the exponent values of the power-law segments $(\Delta s)^{-0.5}$ and 
$(\Delta s)^{-5}$ in Fig~\ref{cumsl}(b) do not give the correct scaling for $\tau$.

\subsection{How the scaling $\tau \sim N^{\alpha}$ comes about}

\begin{figure}
\centerline{
\includegraphics[angle=0, width=0.23\textwidth]{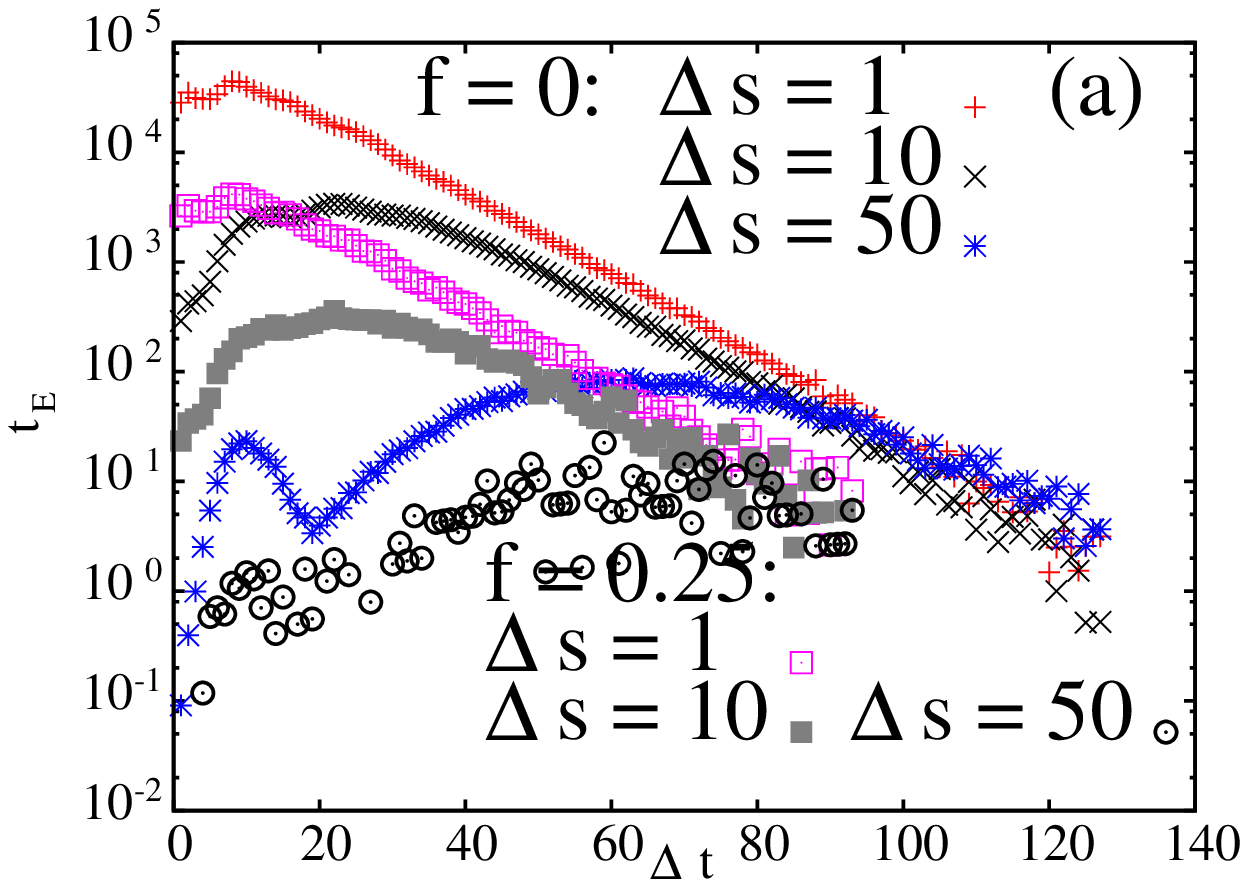}
\includegraphics[angle=0, width=0.23\textwidth]{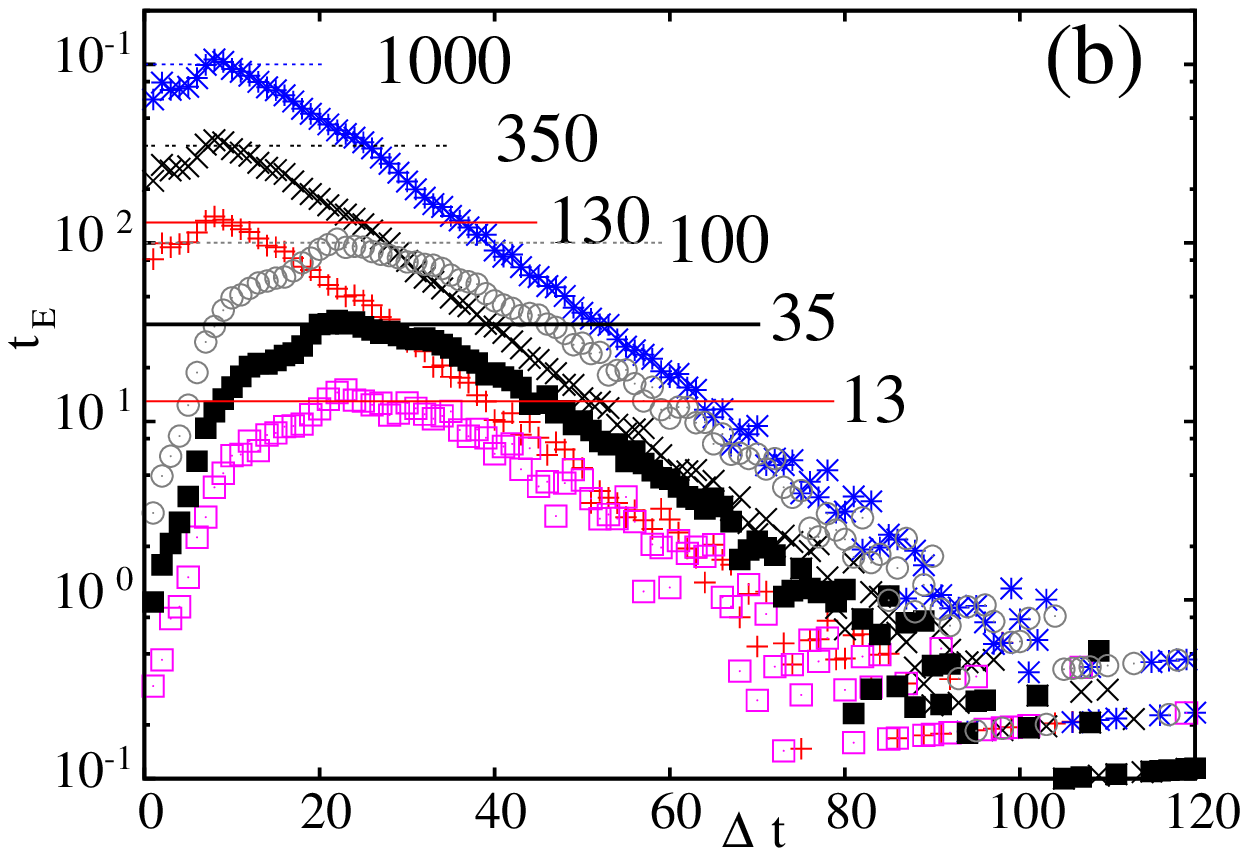}
}
\caption{
(Color online) (a) Distributions of total time $t_E$ for constant $N = 200$ for transferred segment lengths 
$\Delta s \in \{1, 10, 50\}$ and pore force values $f =0$ and $0.25$. (b) The distributions for the constant pore 
force $f = 1$ and different polymer lengths $N$. The bottom three curves with maxima $13$, $35$, and $100$ are for 
$\Delta s = 10$ and $N \in \{50,100,200\}$. The topmost three curves with maxima $130$, $350$, and $1000$ are for 
$\Delta s = 1$ and the same polymer lengths. The marked maxima give the correct scaling $\tau \sim N^\alpha$, see 
the text.
}
\label{sl-f}
\end{figure}

To more precisely determine the role of events at different length scales we extract events of 
three different segment lengths $\Delta s = 1$, $10$, and $50$. Extracting only the transitions of these three 
lengths $\Delta s$ we obtain distributions of the total time $t_E(\Delta s)$ depicting the contribution of 
different individual transition times $\Delta t$. These distributions are shown for pore 
force magnitudes $f = 0$ and $0.25$ in Fig.~\ref{sl-f}(a). The times taken up by events of $\Delta s = 1$ 
are orders of magnitude larger than by events $\Delta s = 10$ or $50$ for all $f \in [0,2]$. 
The peaks in the range $\Delta s \in [0,20]$ most clearly visible for $f = 0$ are due to the contribution of events toward 
{\it cis}. With increasing $f$ this contribution diminishes for $\Delta s = 10$ and $50$ but not for $\Delta s = 1$. This 
partly explains the dominant contribution  of very short transitions to the translocation time. For all $f$, almost all 
the contribution to $\tau$ comes from events of small $\Delta s$ and $\Delta t$. As $\Delta t$ increases, $t_E$ for 
different $\Delta s$ approaches the same value.

\begin{figure}
\includegraphics[angle=0, width=0.4\textwidth]{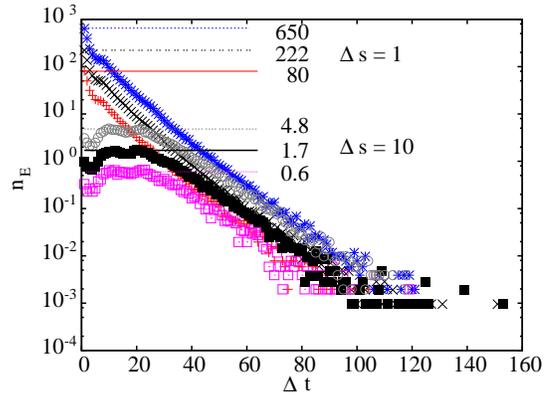}
\caption{
(Color online) The distributions of the numbers $n_E(\Delta s, \Delta t)$ for $\Delta s = 1$, and $10$ for pore force $f = 1$. 
The indicated maxima give $\alpha = 1.5$. 
}
\label{cmpr}
\end{figure}

The forms of the distributions $t_E(\Delta s)$ for constant $\Delta s$ are identical for all $f$, 
including $f = 0$. The maximum values of $t_E$ for $f = 0.25$, $0.5$, and $1$ are 
$\max(t_E) = 360$, $180$, and $105$, respectively, when $\Delta s = 10$, and 
$\max(t_E) = 4200$, $2100$, and $1050$, respectively, when $\Delta s = 1$. These values give the 
scaling $\tau \sim f^{-\beta}$ with $\beta = 1$ for $\Delta s = 1$ and $\beta \lesssim 1$ for 
$\Delta s = 10$.  Fig.~\ref{sl-f}(b) shows that maximum values of $t_E(\Delta s)$ with $\Delta s = 1$ or $10$ give 
the scaling $\tau \sim N^\alpha$, where $\alpha \approx 1.5$. Since the distributions maintain their form, the numbers of 
events $n_E(\Delta t)$ for constant $\Delta s$ should give the same scaling as the total event times $t_E$. In Fig.~\ref{cmpr} 
the event time and 
number distributions for $f = 1$ and $\Delta s \in \{1, 10\}$ are shown. The indicated maxima for the 
number distributions indeed give $\alpha \approx 1.5$ just as the maxima for $t_E$ distributions indicated in 
Fig.~\ref{sl-f}(b).


A more accurate estimate for $\alpha$ can be obtained by summing up all the total event times or total 
event numbers for a constant $\Delta s$.  These give for $\Delta s = 1$: $\alpha = 1.45$, $1.45$, and 
$1.47$ for $f = 0.25$, $0.5$, and $1$, respectively; for $\Delta s = 10$: $\alpha = 1.45$, $1.44$, and 
$1.44$ for $f = 0.25$, $0.5$, and $1$ are obtained, respectively. From the direct measurement of the 
times it takes polymers to translocate we obtain $\alpha = 1.45$, $1.45$, and $1.47$ for $f = 0.25$, 
$0.5$, and $1$, respectively. These are exactly the same values as obtained from the distributions for 
$\Delta s =1$. The statistics for $f = 0.25$ is rather modest, since for such a small pore force a 
vast majority of the simulated polymers slide back to the {\it cis} side without translocating. Even so, exactly 
same estimates for $\alpha$ are obtained by the direct measurement of $\tau$ 
and by summing up all $t_E(\Delta s = 1)$. The number of short 
range and time events determine the polymer translocation even for strongly driven translocation, and 
the effect of mildly changing forms of distributions is negligible.

\section{Conclusion}
\label{cn}

In conclusion, we have characterized the stochastic polymer translocation process with high length 
resolution. Our extensive simulations show that although the unforced and forced translocation 
processes are fundamentally different, the pertinent distributions are almost identical in form. The 
obtained log-normal forms of the event distributions suggest that both processes can be characterized as 
multiplicative stochastic processes. The out-of-equilibrium nature and the fact that the polymer segment transition 
probability through the pore varies with the process state, {\it i.e.} the position of the polymer with 
respect to the pore, also fit the general description of multiplicative processes.

Despite the strong non-equilibrium character of forced polymer translocation, the variation of scaling exponents in 
the relations $\tau \sim N^\alpha$ and $\tau \sim f^{-\beta}$ do not follow from {\it e.g.} the pore force 
changing the pertinent distributions but merely from different numbers of events taking place at a fraction
of the length scale of the model monomer length. Consequently, the noise 
contribution of the thermal heat bath remains significant in spite of the polymer being increasingly lifted out of 
it during the translocation process. The random fluctuations due to the heat bath restrict the lengths of the individual 
transitions of the polymer segments inside the pore. The random stochasticity exerted upon the segments inside the pore 
result from both the random force exerted directly on these segments and the random force resulting from the interaction 
of the dynamically changing polymer conformations and the heat bath outside the pore. Although we have shown that scaling 
of short transition events give the scaling of the whole polymer translocation, the reason for the scaling of the numbers 
of short transition events remains unsolved.

As we have shown previously, the most 
discernible conformational changes are the straightening of the polymer on the {\it cis} side and the crowding of the polymer 
segments on the {\it trans} side~\cite{ourepl,ourpre}. These changes can be described in a coarse-grained fashion even if the
stochasticity is omitted. However, the data presented here confirms that in spite of its strong out-of-equilibrium character 
the polymer translocation is essentially a driven diffusion process when dealing with pore force of realistic 
magnitude. The reported results also show the difficulty of controlling the exact position of the polymer inside the pore. 
Although the average translocation can be described in a coarse-grained manner based on force balance, it will not be of much 
avail in keeping a polymer segment fixed inside the pore. This enhances the importance of the ability to precisely control the 
pore-polymer interaction. This interaction should dominate over the stochastic fluctuations contributing to the random movement 
of the polymer inside the pore if nanopores are to be used for DNA sequencing.
\\

\begin{acknowledgments}
The computational resources of CSC-IT Centre for Science, Finland, are 
acknowledged.
\end{acknowledgments}

\bibliography{pore}

\end{document}